\documentclass[printer]{aa}
\usepackage{natbib}
\usepackage{graphicx}
\usepackage{booktabs}
\usepackage{xspace}
\usepackage{verbatim}
\usepackage{txfonts}

\bibpunct{(}{)}{;}{a}{}{,}

\def\pdot {\dot P}
\def\edot {\dot E}

\def\nh{$N_{\rm H}$\xspace}

\def\flux{erg~s$^{-1}$~cm$^{-2}$\xspace}
\def\lum{erg~s$^{-1}$\xspace}
\def\msun{~M_{\odot}}

\def\deg{^\circ}
\newcommand{\abs}[1]{\vert #1 \vert}
\newcommand{\band}[2]{$#1-#2$ keV\xspace}
\newcommand{\coord}[8]{R.A.\,=\,${#1}^{\rm h} {#2}^{\rm m} {#3}^{\rm s}{#4}$, Dec.\,=\,${#5}^\circ {#6}' {#7}''{#8}$\xspace}

\def\psrj{PSR\,J0726$-$2612\xspace}
\def\xmm{{\em XMM--Newton}\xspace}

\def\cha{{\em Chandra}\xspace}
\def\rosat{{\em ROSAT}\xspace}

\begin{document}

\title{\xmm observations of \psrj, a radio-loud XDINS}

\author{Michela Rigoselli\inst{1,2}, Sandro Mereghetti\inst{1}, Valery Suleimanov\inst{3,4,5}, Alexander Y. Potekhin\inst{6}, Roberto Turolla\inst{7,8}, Roberto Taverna\inst{7,9}, Fabio Pintore\inst{1}}

\institute {INAF, Istituto di Astrofisica Spaziale e Fisica Cosmica Milano, via A.\ Corti 12, I-20133 Milano, Italy
\and
Dipartimento di Fisica G. Occhialini, Universit\`a degli Studi di Milano Bicocca, Piazza della Scienza 3, I-20126 Milano, Italy
\and
Institut fur Astronomie und Astrophysik, Sand 1, 72076 Tubingen, Germany
\and
Kazan (Volga region) Federal University, Kremlevskaja str., 18, Kazan 420008, Russia
\and
Space Research Institute of the Russian Academy of Sciences, Profsoyuznaya Str. 84/32, Moscow 117997, Russia
\and
Ioffe Institute, Politekhnicheskaya 26, 194021, Saint Petersburg, Russia
\and
Dipartimento di Fisica e Astronomia, Universit\`a di Padova, via F. Marzolo 8, I-35131 Padova, Italy
\and
MSSL-UCL, Holmbury St. Mary, Dorking, Surrey RH5 6NT, UK
\and
Dipartimento di Matematica e Fisica, Universit\`a di Roma Tre, via della Vasca Navale 84, I-00146 Roma, Italy
}

\offprints{m.rigoselli@campus.unimib.it, michela.rigoselli@inaf.it}

\date{Received / Accepted}
\date{}

\authorrunning{Rigoselli et al.}

\titlerunning{\psrj}

\abstract{We present the results of an \xmm observation of the slowly rotating ($P=3.4$ s), highly magnetized ($B \approx 3\times10^{13}$ G) radio pulsar \psrj. 
A previous X-ray observation with the \cha satellite showed that some of the properties of \psrj are similar to those of the X-ray Dim Isolated Neutron Stars (XDINSs), a small class of nearby slow pulsars characterized by purely thermal X-ray spectra and undetected in the radio band. 
We confirm the thermal nature of the X-ray emission of \psrj, which can be fit by the sum of two blackbodies with temperatures $kT_1 = 0.074_{-0.011}^{+0.006}$ keV and $kT_2=0.14_{-0.02}^{+0.04}$ keV and emitting radii $R_1=10.4_{-2.8}^{+10.8}$ km and $R_2=0.5_{-0.3}^{+0.9}$ km, respectively (assuming a distance of 1 kpc).
A broad absorption line modeled with a Gaussian profile centred at $0.39_{-0.03}^{+0.02}$ keV is required in the fit.
The pulse profile of \psrj is characterized by two peaks with similar intensity separated by two unequal minima, a shape and pulsed fraction that cannot be reproduced without invoking magnetic beaming of the X-ray emission.  
The presence of a single radio pulse suggests that in \psrj the angles that the dipole axis and the line of sight make with the rotation axis, $\xi$ and $\chi$ respectively, are similar. This geometry differs from that of the two radio-silent XDINSs with a double peaked pulse profile similar to that of \psrj, for which $\xi \sim 90\deg$ and $\chi \sim 45\deg$ have been recently estimated.
These results strengthen the similarity between \psrj and the XDINSs and support the possibility that the lack of radio emission from the latter might simply be due to an unfavourable viewing geometry.

}

\maketitle

\section{Introduction}
Observations with the \rosat satellite in the mid-1990s led to the discovery of a small group of isolated neutron stars characterized by thermal emission at soft X-rays, now known as XDINSs (X-ray Dim Isolated Neutron Stars, see \citealt{hab07, tur09} for reviews). 
XDINSs have spin periods in the range $P\sim 3-17$ s and period derivatives of a few 10$^{-14}$ s s$^{-1}$, which result in characteristic ages $\tau_c = P/2\pdot\sim1-4$ Myr. With the usual assumption that the spin-down is due to magnetic dipole braking, these timing parameters imply magnetic fields of the order of a few $10^{13}$ G.
 
The XDINSs are at distances of only a few hundreds parsecs and for two of them the parallax of the optical counterpart has been measured \citep{wal10,tet11}.
The XDINSs have X-ray luminosities of $10^{31}-10^{32}$ erg s$^{-1}$, higher than their spin-down power. Their X-ray spectra are very soft, with blackbody temperatures of $kT\sim45-110$ eV, often showing the presence of broad absorption lines. If these lines are interpreted as proton cyclotron features or atomic transitions (see, e.g., \citealt{kap08}), the magnetic fields estimated  from their energies are of the same order of those derived from the spin-down rate assuming magnetic dipole braking.
The X-ray emission of XDINSs, consisting  only of thermal components, is believed to come directly from the  star surface and, given the small distance of these sources, it is little affected by photoelectric  absorption in the interstellar medium. The XDINS discovery raised some excitement since they appeared as optimal targets to test neutron star surface emission models without being affected by the presence of non-thermal emission. However, the ultimate goal of constraining the star radius and hence the equation of state with these studies is still hampered by our poor knowledge of the neutron star surface layers composition and magnetization.

\begin{figure}[htbp!]
    \centering
        \includegraphics[width=1\linewidth]{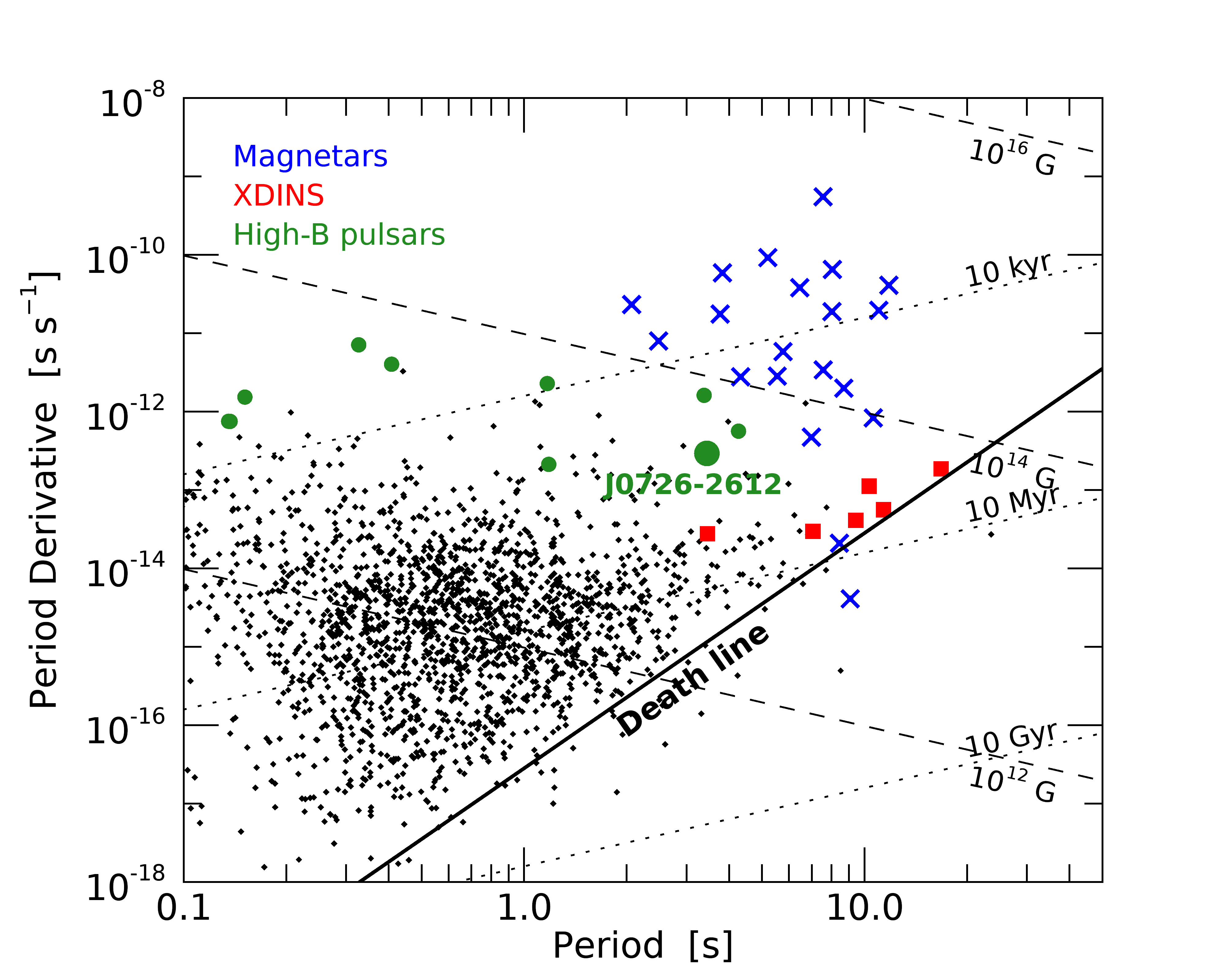}\\
    \caption{$P-\pdot$ diagram of rotation-powered pulsars (black dots) and other classes of isolated pulsars (colored symbols). Lines of equal characteristic age (dotted, $10^{4}-10^{10}$ yr) and equal dipole magnetic field (dashed, $10^{12}-10^{16}$ G) are indicated. The radio pulsar death line $B/P^2 = 1.7 \times 10^{11}$ G s$^{-2}$ \citep{bha92} is also shown. The data are taken from the ATNF Pulsar Catalogue \citep{man05}.}
    \label{fig:p0p1}
\end{figure}

The attempt to explain the different manifestations of neutron stars (e.g. \citealt{mer11}) in the context of a unified evolutionary picture is one of the current challenges in the study of neutron stars \citep{kas10,igo14}. In the $P-\pdot$ diagram, shown in Fig. \ref{fig:p0p1}, XDINSs are located in the region below that occupied by the magnetars, a group of isolated neutron stars powered mainly by magnetic energy (see, e.g. \citealt{mer15,tur15,kas17}).  
This has led to the suggestion that the XDINSs might be the descendent of magnetars \citep{hey98,col00}. The strong internal field of magnetars ($B\gtrsim10^{15}$ G) significantly affects their thermal evolution \citep{vig13}, resulting in luminosities higher than those predicted for normal pulsars of similar age. 

A distinctive property of the XDINSs is that they are not detected in the radio band\footnote{The possible detection of pulsed emission from two XDINSs at very low frequencies \citep{mal05,mal06}  is, so far, unconfirmed.} \citep{kon09}. 
The reason for the lack of radio emission is still uncertain. One possibility is that this is due  to their old age and long spin period \citep{bar98,bar01}. 
However, a few radio pulsars with periods $\gtrsim\!10$ s have been recently discovered: PSR J0250$+$5854 with $P=23.5$ s \citep{tan18}, and a second one with $P=12.1$ s (Morello et al. 2019 in prep). Another explanation might be related to the geometrical configuration of their magnetosphere, that, especially if they are old magnetars, might be strongly non-dipolar \citep{tur15}.
Finally, it cannot be excluded that (at least some of) the XDINSs are simply ordinary radio pulsars with radio beams unfavorably aligned with respect to the Earth. 
In this respect, it is interesting to investigate radio-loud pulsars with X-ray properties and/or timing parameters similar to those of the XDINSs, such as the long period (greater than a few seconds) and high B ($>10^{13}$ G) pulsars. 

Among these, here we focus on \psrj, a radio pulsar with spin period $P=3.44$~s and characteristic age of 200 kyr that was discovered in the Parkes High-Latitude Survey \citep{bur06}. Its timing parameters (Table \ref{tab:prop}) are in the range of those of the XDINSs. 
The similarity with the XDINSs was reinforced by X-ray observations with the \cha satellite \citep{spe11}, that revealed a soft thermal spectrum with blackbody temperature $kT\approx87$ eV, and pulsations with a sinusoidal, double-peaked  profile. 
The distance of \psrj is unknown. Its dispersion measure DM $=69.4 \pm 0.4$ cm$^{-3}$ pc \citep{bur06} implies a distance $d=2.9$ kpc, assuming the Galactic electrons distribution of \citet{yao17}. However, there are a few facts suggesting that this is probably an overestimate. Such a large value would give a distance of 230 pc from the Galactic plane, implying, if \psrj was born close to the plane and its true age is similar to $\tau_c$, a velocity of the order of a thousand km s$^{-1}$. This value is not impossible, but it would be at the far end of the pulsar velocity distribution \citep{hob05}.  More importantly, for such a large $d$, one would expect an X-ray absorption corresponding to a sizeable fraction of the total Galactic H I column density, that in this direction is $\sim5\times10^{21}$ cm$^{-1}$ \citep{kal05}, while the observed value is a factor 10 smaller. 
Finally, the line of sight toward \psrj crosses the Gould belt, that is not included in the electron distribution model of \citet{yao17}. This could explain the large distance inferred from the DM. This local structure ($d\sim200-400$ pc) comprises several OB associations that have been proposed as the birthplace of the XDINSs \citep{pop03,pop05}. \citet{spe11} suggested that also \psrj could be associated with the Gould belt and hence closer than $\sim1$ kpc.

Here we report the results of \xmm observations which show other similarities between \psrj and the XDINSs. In the following we will scale all the distance-dependent quantities to $d_{\rm kpc}=1$ kpc and adopt representative values of mass and radius of $1.2\msun$ and 12 km, respectively.

\section{Observations and data reduction}

\begin{table}[htbp!]
\centering \caption{Observed and derived parameters for \psrj}
\label{tab:prop}

\begin{tabular}{ll}
\toprule
R.A. (J2000.0)\dotfill										& $07^\mathrm{h}26^\mathrm{m}08^\mathrm{s}.12(4)$ \\
Dec. (J2000.0)\dotfill										& $-26^\circ12'38''.1(8)$ \\
Period $P$ (s)\dotfill										& $3.4423084877(4)$\\
Period derivative $\pdot$ (s s$^{-1}$)\dotfill				& $2.9311(4)\times 10^{-13}$ \\
Epoch (MJD)\dotfill											& $52,950$\\
Characteristic age $\tau_c$ (years)\dotfill					& $1.86\times 10^5$ \\
Surface dipolar magnetic field $B_s$ (G)\dotfill			& $3.2\times 10^{13}$ \\
Rotational energy loss rate $\edot$ (erg s$^{-1}$)\dotfill	& $2.8\times 10^{32}$ \\
Dispersion measure DM (cm$^{-3}$ pc)\dotfill				& $69.4(4)$\\
\bottomrule\\[-5pt]
\end{tabular}

\raggedright
Data are taken from \citet{bur06} and the ATNF Pulsar Catalogue \citep{man05}. Numbers in parentheses show the $1\sigma$ uncertainty for the last digits. \\
\end{table}

\psrj was observed with the European Photon Imaging Cameras (EPIC) instrument on board \xmm with a single pointing lasting 108 ks on 2013 April 8.
The three cameras of EPIC (\band{0.1}{12}), the pn \citep{str01} and the two MOS \citep{tur01}, were operated in Full Frame mode with the thin optical filter. While the pn time resolution (73.4 ms) is adequate to reveal the pulsations of the source, this is impossible for the MOS given its resolution time of 2.6 s. 

The data reduction was performed using the \textsc{epproc} and \textsc{emproc} pipelines of version 15 of the Science Analysis System (SAS)\footnote{https://www.cosmos.esa.int/web/xmm-newton/sas}.
We selected single- and multiple-pixel events (PATTERN $\leq4$ and PATTERN $\leq12$) for both the pn and MOS.
We then removed time intervals of high background using the SAS program \textsc{espfilt} with standard parameters. The source was detected by EPIC at coordinates \coord{07}{26}{08}{.1}{-26}{12}{38}{}, fully consistent with the radio position (Table \ref{tab:prop}). 
The source events were selected from a circle of radius $40''$ centred at the radio position, while the background was extracted from a nearby circular region of radius $60''$. The resulting net exposure times and source events are listed in Table \ref{tab:counts}. At the corresponding count rates pile-up effects are not relevant. 

\begin{table}[htbp!]
\centering \caption{Exposure Times and Source Counts for \psrj in the three EPIC cameras}
\label{tab:counts}
\begin{tabular}{lcccc}
\toprule
Data            & EPIC camera	& Exposure time	& Source Counts	        \\
		        &	   	        & ks		    & \band{0.15}{1.5}  \\
\midrule
Phase-averaged	& pn	    	& 37.8	    	& $18,938 \pm 140$   \\
		        & MOS1	    	& 64.0	    	& $4,499 \pm 69$     \\
		        & MOS2	    	& 70.4	    	& $5,212 \pm 74$     \\
\midrule
Min 1		    & pn	    	& 9.4		    & $3,823 \pm 63$      \\
Max 1		    & pn	    	& 9.4	    	& $5,576 \pm 76$      \\
Min 2		    & pn		    & 9.4		    & $4,088 \pm 65$      \\
Max 2		    & pn		    & 9.4		    & $5,447 \pm 75$      \\
\bottomrule
\end{tabular}
\end{table}

\section{Results}
\subsection{Timing analysis}
\begin{figure}[htbp!]
    \centering
        \includegraphics[trim=0cm 1cm 0cm 14cm,clip,width=1\linewidth]{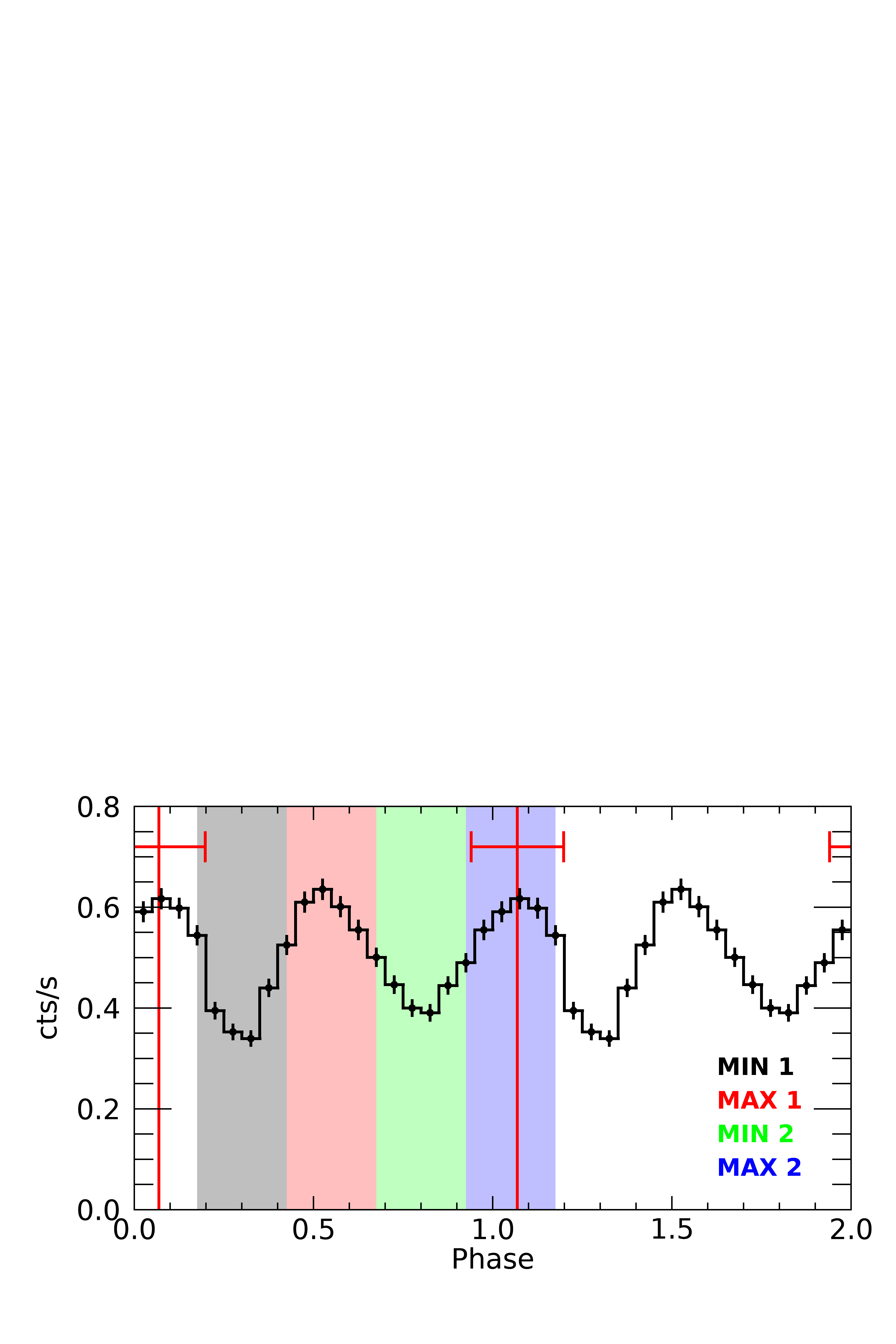}\\
    \caption{Pulse profile of \psrj in the energy range \band{0.15}{1.5} obtained by folding the EPIC-pn data in 20 phase bins at the period derived from the radio ephemeris (Table \ref{tab:prop}). The vertical red line represents the location of the radio pulse (derived from \citealt{spe11}), with its uncertainty ($1\sigma$). The colored bands indicate the intervals used for the phase-resolved spectroscopy.}
    \label{fig:lcall}
\end{figure}

\psrj is barely detected above 1.5 keV, therefore we limited our timing analysis to the energy band \band{0.15}{1.5}.
The times of arrival were converted to the barycenter of the Solar System with the task \textsc{barycen}.
An epoch folding search of the EPIC-pn data gave a best period $P = 3.442396(1)$ s, that is consistent within $0.7\sigma$ with the value expected at the \xmm observation epoch ($56,390$ MJD) using the ATNF ephemeris reported in Table \ref{tab:prop}. The background-subtracted light curve, in the energy band \band{0.15}{1.5}  is shown in Fig.~\ref{fig:lcall}.
The position of the radio pulse is indicated, with its 1$\sigma$ uncertainty, as a vertical red line.

The EPIC-pn pulse profile shows two peaks with the same intensity (net count rate of max$_1 = 0.62 \pm 0.02$ cts s$^{-1}$ and max$_2 = 0.64\pm0.02$ cts s$^{-1}$), separated by about 0.5 cycles. The two minima of the pulse profile are instead significantly different: min$_1 = 0.34 \pm 0.01$ cts s$^{-1}$ and min$_2 = 0.39 \pm 0.01$ cts s$^{-1}$. 
The pulse profile is symmetric in phase with respect to any of the two minima, but a fit with a constant plus a sine function at half of the spin period is not acceptable ($\chi^2_\nu=2.7$ for 17 dof).
The pulsed fraction\footnote{Defined as (max(CR)-min(CR))/(max(CR)+min(CR)), where CR is the background-subtracted count rate.} is $30\pm2\%$.

\begin{figure}[htbp!]
    \centering
    \includegraphics[width=1\linewidth]{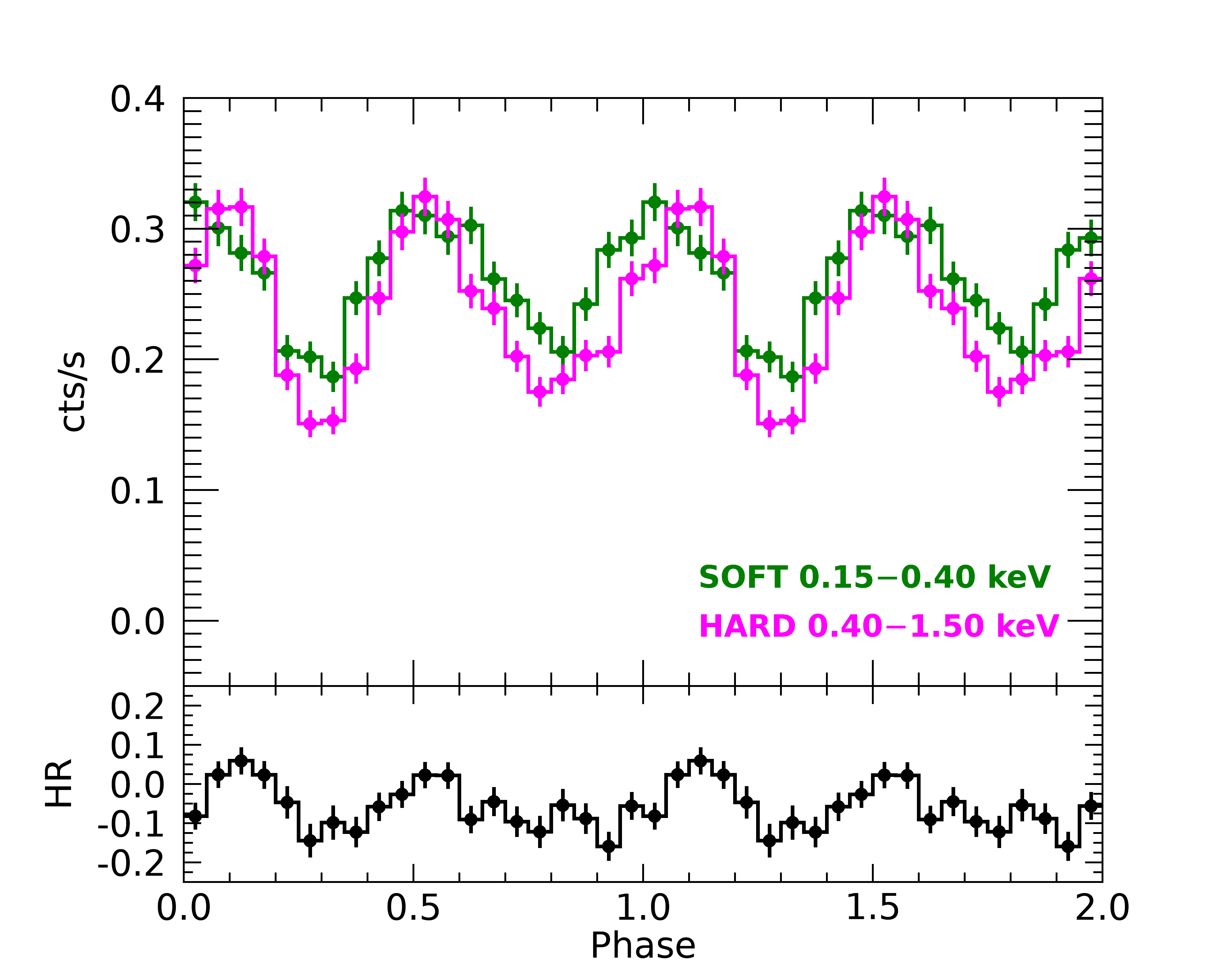}
    \caption{EPIC-pn light curve of \psrj (20 phase bins) in the energy ranges $0.15-0.4-1.5$ keV together with the corresponding hardness-ratio.}
    \label{fig:lcband}
\end{figure}

Fig. \ref{fig:lcband} shows that the soft ($0.15-0.4$ keV) and hard  energy ranges ($0.4-1.5$ keV) have slightly different pulsed fractions: $26\pm3\%$ and $37\pm3\%$, respectively. Moreover, the positions of the first minimum and of the second maximum are shifted of about 1 bin between the two energy ranges, but the symmetry around the minima is preserved in both bands. 
Fits with a constant plus sine function give $\chi^2_\nu=1.6$ and $\chi^2_\nu=3.5$ for the soft and hard profile, respectively.
The hardness ratio\footnote{Defined as (hard(CR)-soft(CR))/(hard(CR)+soft(CR)), where the soft energy range is \band{0.15}{0.4}, the hard one \band{0.4}{1.5}.}, shown in the lower panel of the same figure, clearly indicates the presence of phase-dependent spectral variations: the source is softer during the minima and harder during the maxima.

\subsection{Spectral analysis} \label{sec:spec}

\begin{figure}[htbp!]
    \centering
    \includegraphics[width=0.9\linewidth]{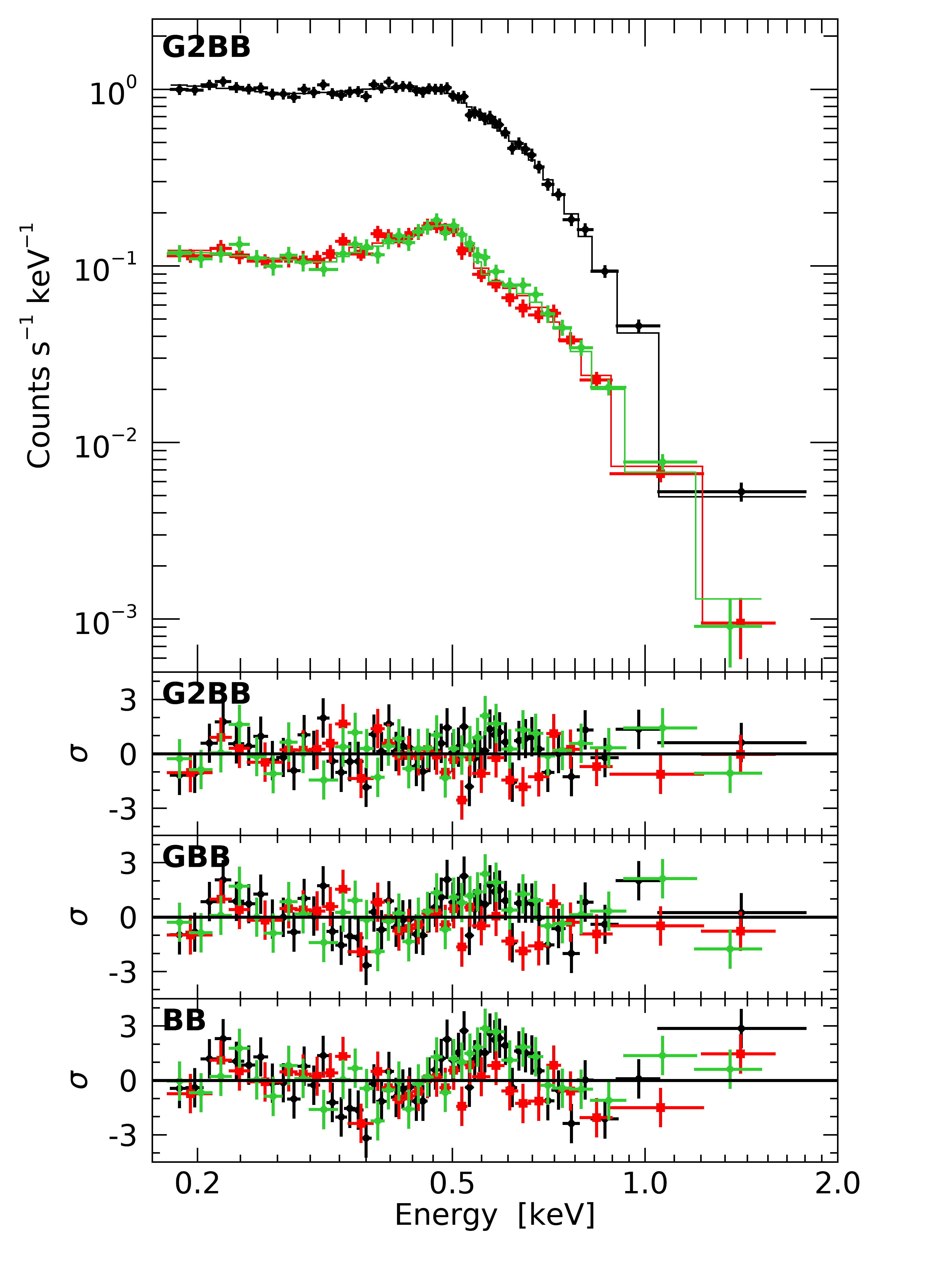}
    \caption{EPIC-pn (black), -MOS1 (red) and -MOS2 (green)  phase-averaged spectra of \psrj. The top panel shows the best fit using a Gaussian absorption feature at $E=0.39$ keV and two blackbodies (G2BB). The lower panels show the residuals of the best fit (G2BB), of a Gaussian absorption feature at $E=1.09$ keV and one blackbody (GBB) and of a single blackbody (BB) in units of $\sigma$. Data have been rebinned for display purposes only.}
    \label{fig:spec}
\end{figure}

The spectral analysis was performed using XSPEC (ver. 12.8.2). The spectra were rebinned using the GRPPHA tool with a minimum of 50 counts per bin. The spectra of the three cameras were fitted simultaneously, including a renormalization factor to account for possible cross-calibration uncertainties.
Errors on the spectral parameters are at $1\sigma$ confidence level.

We used the photoelectric absorption model \textsc{tbabs}, with cross sections and abundances from \citet{wil00}.
Both a single power law and a blackbody did not provide acceptable fits, giving $\chi^2_\nu \approx 6$ and $\chi^2_\nu = 1.37$ for 213 dof (Null Hypothesis Probability, nhp, of $3\times 10^{-4}$), respectively.
We then attempted a fit with magnetized hydrogen atmosphere models (\textsc{nsa} and \textsc{nsmaxg} in XSPEC, \citealt{pav95,ho08,ho14}). However, none of the two sets of available models (the first with a single surface $B$ and $T_{\rm eff}$, the second with $B$ and $T_{\rm eff}$ varying across the surface according to the magnetic dipole model) gave an acceptable fit ($\chi^2_\nu > 2.2$ for 213 dof).
In conclusion, we could not find a good fit with single component models.

Also modelling the spectra with a blackbody plus power law or with the sum of two blackbodies was unsatisfactory. In the first case we obtained a negative photon index for the power law, while in the second case, the second thermal component had a negligible flux, and did not improve the quality of the fit with respect to that of a single blackbody ($\chi^2_\nu = 1.32$ for 211 dof, nhp $=10^{-3}$).

A real improvement in the fit was obtained by adding to the blackbody a broad absorption line modelled with a Gaussian (GBB) centered at $E=1.09 \pm 0.09$ keV and width $\sigma=0.28\pm0.08$ keV ($\chi^2_\nu = 1.12$ for 210 dof). Following the recent results of \citet{yon19}, we explored the possibility to adopt a two blackbody component model plus a Gaussian line in absorption (G2BB). With this model we found a good fit with the line placed at $E=0.39_{-0.03}^{+0.02}$ keV and with a broadening of $\sigma=0.08_{-0.02}^{+0.03}$ keV ($\chi^2_\nu = 1.00$ for 208 dof). The addition of the line yields an improvement of the $\chi^2$ of $F=\chi^2_{\rm 2BB}/\chi^2_{\rm G2BB}=1.32$. To assess the statistical significance of the line, we estimated through Monte Carlo simulations the probability of obtaining by chance an equal (or better) fit improvement: we estimate a probability of $\sim 10^{-5}$ of having $F \geq1.32$, corresponding to a $\sim 4.4\sigma$ significance of the line. The cold blackbody ($kT_1 \approx 0.074$ keV) has an emitting radius $R_1=10.4_{-2.8}^{+10.8}$ $d_{\rm kpc}$ km, compatible with emission from the whole neutron star, while the hot blackbody has $kT_2 \approx 0.14$ keV and $R_2=0.5_{-0.3}^{+0.9}$ $d_{\rm kpc}$ km. 

A good fit was also found with the magnetized atmosphere models with a dipole distribution of the surface magnetic field ($B=10^{13}$ G at the poles) plus a Gaussian line in absorption. With the \textsc{nsa} model, we found an effective temperature $T_{\rm eff} = 0.40\pm0.08$ MK (corresponding to an observed temperature $kT=0.029\pm0.001$ keV), $d=121_{-12}^{+13}$ pc and $E=0.37_{-0.03}^{+0.02}$ keV, $\sigma=0.09_{-0.01}^{+0.02}$ keV for the Gaussian line ($\chi^2_\nu = 1.03$ for 210 dof). With the \textsc{nsmaxg} model, for an impact parameter (that is the angle between the line of sight and the dipole axis) $\eta=90\deg$, the model parameters are $T_{\rm eff} = 0.39\pm0.02$ MK ($kT=0.028\pm0.001$ keV), $d=63_{-17}^{+26}$ pc and $E=0.28\pm0.09$ keV, $\sigma=0.14_{-0.04}^{+0.06}$ keV for the Gaussian line ($\chi^2_\nu = 1.02$ for 210 dof). Using instead the same model with $\eta=0\deg$, the fit was not acceptable ($\chi^2_\nu = 2.38$ for 210 dof).

The spectral results are summarized in Table \ref{tab:spec}, while in Fig. \ref{fig:spec} the best blackbody fits are shown.

\setlength{\tabcolsep}{0.17em}
\begin{table*}[htbp!]
\centering \caption{Results for the phase-averaged and phase-resolved spectra of \psrj}
\label{tab:spec}

\begin{tabular}{lccccccccccc}
\toprule
Model				& \nh$^{\rm a}$			      & $kT_1$	& $R_1^{\rm~b}$	& $kT_2$    & $R_2^{\rm~b}$ & $E$		& $\sigma$	& strength$^{\rm~c}$ & $F^{0.1-2}_{\rm unabs}$		& $\chi_\nu^2$/dof  & nhp\\[5pt]
					& $10^{20}$ cm$^{-2}$ & keV 	& km 			& keV		& km 		    & keV		& keV 		& keV   & \flux  		& 			        & \\[5pt]
\midrule\\[-5pt]
\multicolumn{11}{l}{Phase-averaged spectra:}\\[5pt]

BB			& $4.1\pm0.2$	& $0.0896(6)$	& $4.90\pm0.15$	& \dots		& \dots		& \dots		& \dots		& \dots		& $1.60_{-0.05}^{+0.06}$	& 1.37/213		& $3\times 10^{-4}$\\[5pt]

2BB			& $4.3\pm0.2$	& $0.0888(7)$	& $5.1\pm0.2$	& $>0.33$	& $<0.018$& \dots		& \dots		& \dots		& $1.65_{-0.06}^{+0.07}$	& 1.32/211		& $1\times 10^{-3}$\\[5pt]

GBB		& $2.8\pm0.3$	& $0.11\pm0.01$	& $2.9_{-0.4}^{+0.5}$	& \dots		& \dots		& $1.09\pm0.09$	& $0.28\pm0.08$	& $1.0_{-0.6}^{+1.1}$	& $1.37_{-0.09}^{+0.35}$	& 1.12/210 	& $0.11$\\[5pt]

G2BB	& $5.3_{-0.8}^{+1.2}$	& $0.074_{-0.011}^{+0.006}$	& $10.4_{-2.8}^{+10.8}$& $0.14_{-0.02}^{+0.04}$	& $0.5_{-0.3}^{+0.9}$ & $0.39_{-0.03}^{+0.02}$	& $0.08_{-0.02}^{+0.03}$	& $0.12_{-0.05}^{+0.13}$	& $3.30_{-0.85}^{+3.85}$	& 1.00/208 	& $0.47$\\[5pt]

GNSA$^{\rm~e}$	& $6.9_{-1.1}^{+0.8}$	& $0.029(1)$	& $14.3^{\rm d}$	& \dots		& \dots		& $0.37_{-0.03}^{+0.02}$	& $0.09_{-0.01}^{+0.02}$	& $0.17_{-0.04}^{+0.07}$	& $9.0 \pm 1.1$	& 1.03/210 	& $0.36$\\[5pt]

GNSMAXG$^{\rm~f}$	& $5.9_{-4.2}^{+3.4}$	& $0.028(1)$	& $14.3^{\rm d}$	& \dots		& \dots		& $0.28\pm0.09$	& $0.14_{-0.04}^{+0.06}$	& $0.62_{-0.33}^{+1.26}$	& $19.3_{-7.0}^{+9.7}$	& 1.02/210 	& $0.40$\\[5pt]

\midrule\\[-5pt]
\multicolumn{11}{l}{G2BB phase-resolved:}\\[5pt]

Maxima 1	& $5.3^{\rm d}$	& $0.074^{\rm d}$	& $10.4^{\rm d}$	& $0.110(7)$		& $1.55_{-0.30}^{+0.40}$  & $0.39\pm0.01$			& $0.07 \pm 0.01$		& $0.11_{-0.01}^{+0.02}$ & $3.6\pm0.2$	& 1.00/80 	& $0.47$	\\[5pt]

Maxima 2	& $5.3^{\rm d}$	& $0.074^{\rm d}$	& $10.4^{\rm d}$	& $0.111(9)$		& $1.40_{-0.35}^{+0.45}$   &$0.39\pm0.01$			& $0.06 \pm 0.01$		& $0.10\pm0.01$			& $3.6\pm0.2$	& 0.95/78 	& $0.61$	\\[5pt]

Minima 1	& $5.3^{\rm d}$	& $0.074^{\rm d}$	& $10.4^{\rm d}$	& $0.17_{-0.04}^{+0.07}$& $0.20_{-0.13}^{+0.31}$& $0.40_{-0.02}^{+0.01}$& $0.14_{-0.02}^{+0.04}$& $0.23_{-0.03}^{+0.05}$ & $2.90_{-0.25}^{+0.45}$	& 1.14/49 	& $0.23$	\\[5pt]

Minima 2	& $5.3^{\rm d}$	& $0.074^{\rm d}$	& $10.4^{\rm d}$	& $0.29_{-0.09}^{+0.72}$& $0.06_{-0.03}^{+0.05}$	& $0.39_{-0.03}^{+0.02}$& $0.13_{-0.02}^{+0.03}$& $0.20_{-0.02}^{+0.04}$ & $3.1_{-0.2}^{+0.3}$	& 1.28/52 	& $0.08$	\\[5pt]

\bottomrule\\[-5pt]
\end{tabular}

\raggedright
Joint fits of EPIC-pn+MOS1+MOS2 phase-averaged spectra and EPIC-pn phase-resolved spectra of \psrj. The fluxes, corrected for the absorption, are expressed in units of $10^{-12}$~erg~cm$^{-2}$~s$^{-1}$. Temperatures and radii are observed quantities at infinity.
Errors at $1\sigma$.\\
$^{\rm a}$ Derived with the photoelectric absorption model \textsc{tbabs} \citep{wil00}.\\
$^{\rm b}$ Radius for an assumed distance of 1 kpc.\\
$^{\rm c}$ Parameter of \textsc{gabs} model such as the optical depth at line center is $\tau=\mathrm{strength}/\!\sqrt{2\pi}\sigma$.\\
$^{\rm d}$ Fixed value.\\
$^{\rm e}$ \textsc{nsa} model \citep{pav95} with $M=1.2\msun$, $R=12$ km, $B=10^{13}$ G and a uniform temperature distribution. This model yields a best fit distance $d=121_{-12}^{+13}$ pc.\\
$^{\rm f}$ \textsc{nsmaxg} model \citep{ho08,ho14} with $M=1.2\msun$, $R=12$ km, a dipole distribution of the magnetic field ($B=10^{13}$ G at the poles) and consistent temperature distribution, seen with $\eta=90\deg$. This model yields a best fit distance $d=63_{-17}^{+26}$ pc.\\

\end{table*}

\begin{figure}[htbp!]
    \centering
    \includegraphics[width=0.9\linewidth]{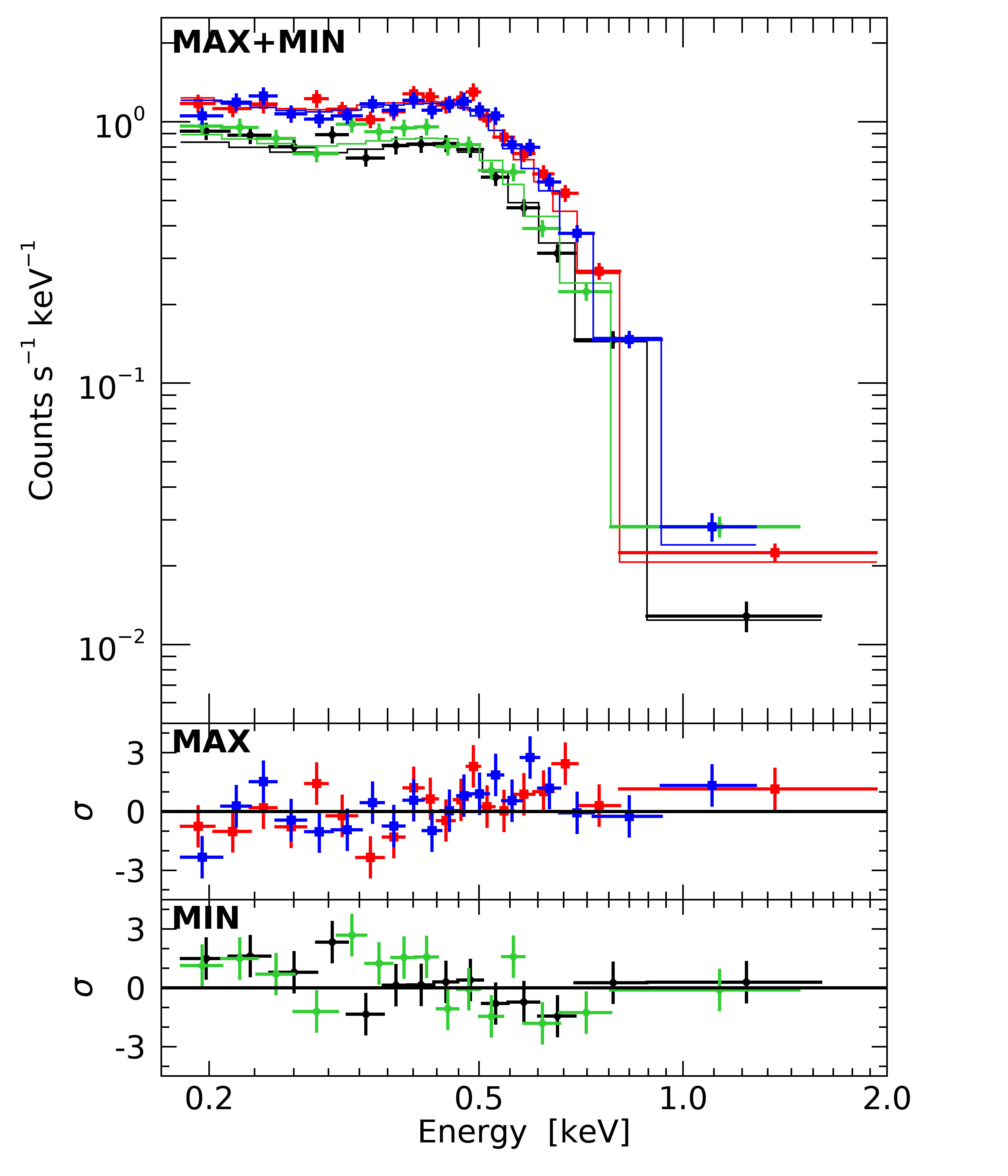}
    \caption{EPIC-pn phase-resolved spectra fitted with the G2BB model used for the phase-averaged spectra (the color code is the same as in Fig. \ref{fig:lcall}). The overall normalization is the only free parameter. The residuals of the spectra at maxima and minima, in units of $\sigma$, are shown in the lower panels. Data have been rebinned for display purposes only.}
    \label{fig:specres}
\end{figure}

The light curves and hardness ratio shown in Fig. \ref{fig:lcband} indicate that a spectral variation occurs as a function of the rotation phase. Therefore, we extracted the EPIC-pn spectra of the phase intervals corresponding to the two minima and the two maxima of the pulse profile, as shown in Fig. \ref{fig:lcall} (the number of source events in each spectrum is listed in Table \ref{tab:counts}).
In order to illustrate the spectral variations, we fitted the spectra with the G2BB model, fixing all of the parameters at the best fit values of the phase-averaged spectrum, except for an overall normalization. The residuals, shown in the two lower panels of Fig.~\ref{fig:specres}, indicate that the spectra of the two maxima are similar and significantly harder then those of the minima. Their normalization factors with respect to the phase-averaged spectrum, $N_{\rm max1} = 1.16\pm0.02$ and $N_{\rm max2}=1.14\pm0.02$, are consistent with the same value, while those of the two minima are different ($N_{\rm min1} = 0.85\pm0.01$ and $N_{\rm min2}=0.80\pm0.01$).

We then fitted the four spectra separately, keeping fixed only the interstellar absorption and the parameters of the cold blackbody, because we do not expect them to vary during a stellar rotation. 
The results are given in Table \ref{tab:spec}. The absorption line is at the same energy in the four spectra, but it has different widths and normalizations. The hot blackbody temperature is lower ($kT \approx 0.11$ keV) and its emission radius is larger ($R \approx 1.5$ $d_{\rm kpc}$ km) at the two maximum phases than at the first minimum ($kT \approx 0.17$ keV and $R \approx 0.20$ $d_{\rm kpc}$ km), while these parameters are poorly constrained at the second minimum.
We also tried other fits allowing more parameters to vary, but the results were inconclusive due to the strong parameter degeneracy.

\section{Discussion}

Our \xmm results for \psrj are consistent with those  previously obtained with \cha \citep{spe11}, but, thanks to a significant detection with good statistics over a broader energy range, they provide more information on the spectrum and pulse profile of this pulsar.

We found that the spectrum of \psrj is more complex than the single blackbody that was adequate to fit the \cha data. 
The single blackbody fit requires the addition of a broad absorption line at $E\approx1.09$ keV. 
A better fit was obtained with two blackbody components, but also in this case a line at $E\approx0.39$ keV is required. 
The colder blackbody component has an emitting area consistent with a large fraction of the star surface ($R_1=10.4_{-2.8}^{+10.8}$ $d_{\rm kpc}$ km), while the hotter one can be attributed to a small hot spot ($R_2=0.5_{-0.3}^{+0.9}$ $d_{\rm kpc}$ km), likely located at the magnetic pole.

Our results confirmed that the interstellar absorption is about a factor of ten smaller than the value (\nh $=2.1 \times 10^{21}$ cm$^{-2}$) inferred from the dispersion measure and the usual assumption of a $10\%$ ionization of the interstellar medium \citep{he13}. This might be due to the line of sight crossing the Gould belt.

An equally good fit was obtained with a magnetized hydrogen atmosphere covering all the star surface, but also in this case the presence of an absorption line at $E\approx0.37$ keV (\textsc{nsa} model) or $E\approx0.28$ keV (\textsc{nsmaxg} model) is required. We note that the constant (polar) value of the magnetic field in the \textsc{nsa} (\textsc{nsmaxg}) model is fixed in the fits at $B=10^{13}$ G, and that the \textsc{nsa} model assumes a uniform distribution of the temperature. The \textsc{nsmaxg} model is more realistic, but it assumes that the dipole axis is orthogonal to the line of sight, that does not necessarily apply to the case of \psrj. Moreover, the inferred distance of $\approx 63$ pc seems unrealistically small.

The absorption lines we found in the spectra can be interpreted as proton cyclotron features at $E_{\rm cyc} = 0.063~ B_{13} \times (1+z)$ keV, where $z$ is the gravitational redshift and $B_{13}$ the  magnetic field in units of $10^{13}$ G.
In the case of G2BB model, for $E_{\rm cyc}=0.39$ keV and $z\approx0.2$, we get $B\approx 5\times 10^{13}$ G, in good agreement with the dipole magnetic field evaluated at the poles ($B_p\approx 6\times 10^{13}$ G). 
However, we caution that other explanations cannot be ruled out, including the possibility that the lines are simply an artefact resulting from an oversimplified  modeling of the continuum emission. In fact, \citet{vig14} showed that non-homogeneus temperature distributions on a neutron star surface can, in some cases, lead to the appearance of broad features when the spectra are fitted with simple blackbody models.

Contrary to the previous \cha results, we also found that  the double-peaked pulse profile of \psrj is not  well  described by a sinusoid, owing to the significant difference in the flux of the two minima.
Remarkably, the pulse profile is symmetric for phase reflection around any of the two minima. Within the limits due to their lower statistics, these properties seem to hold also for the profiles in the soft and hard X-ray bands. The pulse profiles are moderately energy-dependent, with evidence for a harder emission in correspondence of the two peaks.

\begin{figure}[htbp!]
    \centering
     \includegraphics[width=1\linewidth]{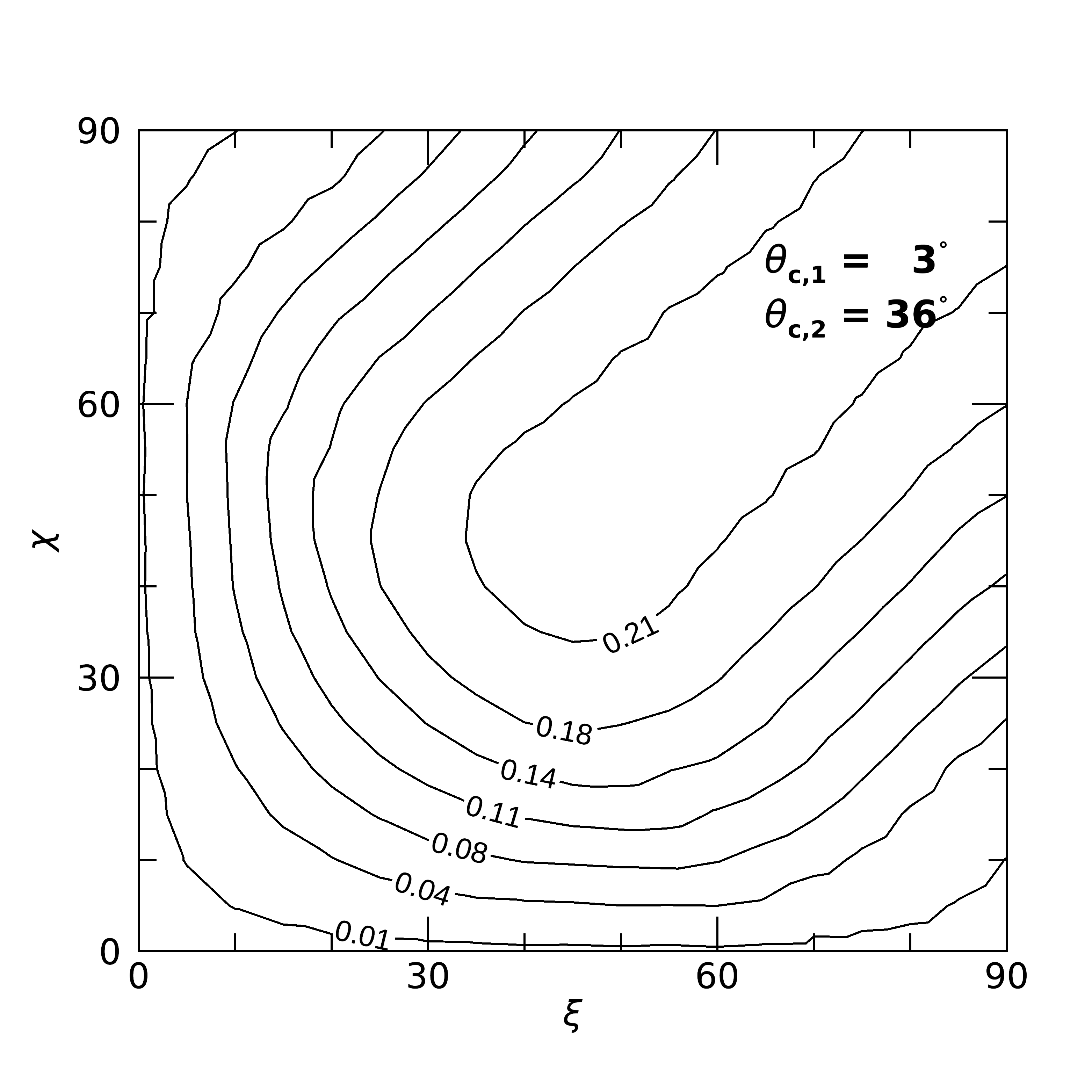}
    \caption{Pulsed fraction for the G2BB model, where the blackbody emission comes from two, antipodal ``cap+ring'' spots centred on the magnetic poles. The aperture of the hot cap ($kT=0.14$ keV) is $\theta_{c,1}=3\deg$, while the colder ($kT=0.074$ keV) ring extends from $\theta_{c,1}$ to $\theta_{c,2}=36\deg$. The considered energy range is \band{0.15}{1.5}, and a compactness of $M/R=0.1\msun/$km has been assumed.}
    \label{fig:pfchixi}
\end{figure}

Although a detailed modeling of the light curves of \psrj is beyond the scope of the present work,  we  explored whether a simple model based on blackbody emission components with parameters consistent with the spectral results could reproduce the pulse profile. 
We assumed that the hotter blackbody comes from two antipodal magnetic polar caps with opening angle $\theta_{c1}=3^{\circ}$, while the colder one from two annuli extending between $\theta_{c1}$ and $\theta_{c2}=36\deg$.  
The temperatures of the emitting regions were set to the values derived from the spectral analysis (model G2BB, $kT_1=0.074$ keV, $kT_2=0.14$ keV) and the angular apertures were chosen in such a way to reproduce the emitting radii derived from the fit for a NS radius of 12 km. We also added interstellar absorption and a Gaussian absorption line, with parameters fixed to those of the phase averaged spectrum. Synthetic light curves were computed using the method by \citet{tur13} and account for general-relativistic effects. We convolved the obtained light curves with the EPIC-pn instrumental response and we evaluated the pulsed fraction in the energy range \band{0.15}{1.5}. 
The results depend on the  angles $\chi$ and $\xi$ that the rotation axis makes with the line of sight and the magnetic axis, respectively.
As shown in Fig. \ref{fig:pfchixi}, this simple model is unable to yield the observed pulsed fraction even for the most favourable geometry (PF $\approx21\%$ for $\xi\approx\chi\gtrsim 35\deg$). This is also true if only two antipodal point-like polar caps are considered, which is the configuration yielding the maximum pulsed fraction using isotropic emission (see e.g. \citealt{tur13}). Another problem is that, owing to the intrinsic symmetry of the model, the resulting light curves cannot exhibit different minima, as observed in \psrj. 

\begin{figure}[htbp!]
    \centering
    \includegraphics[width=1\linewidth]{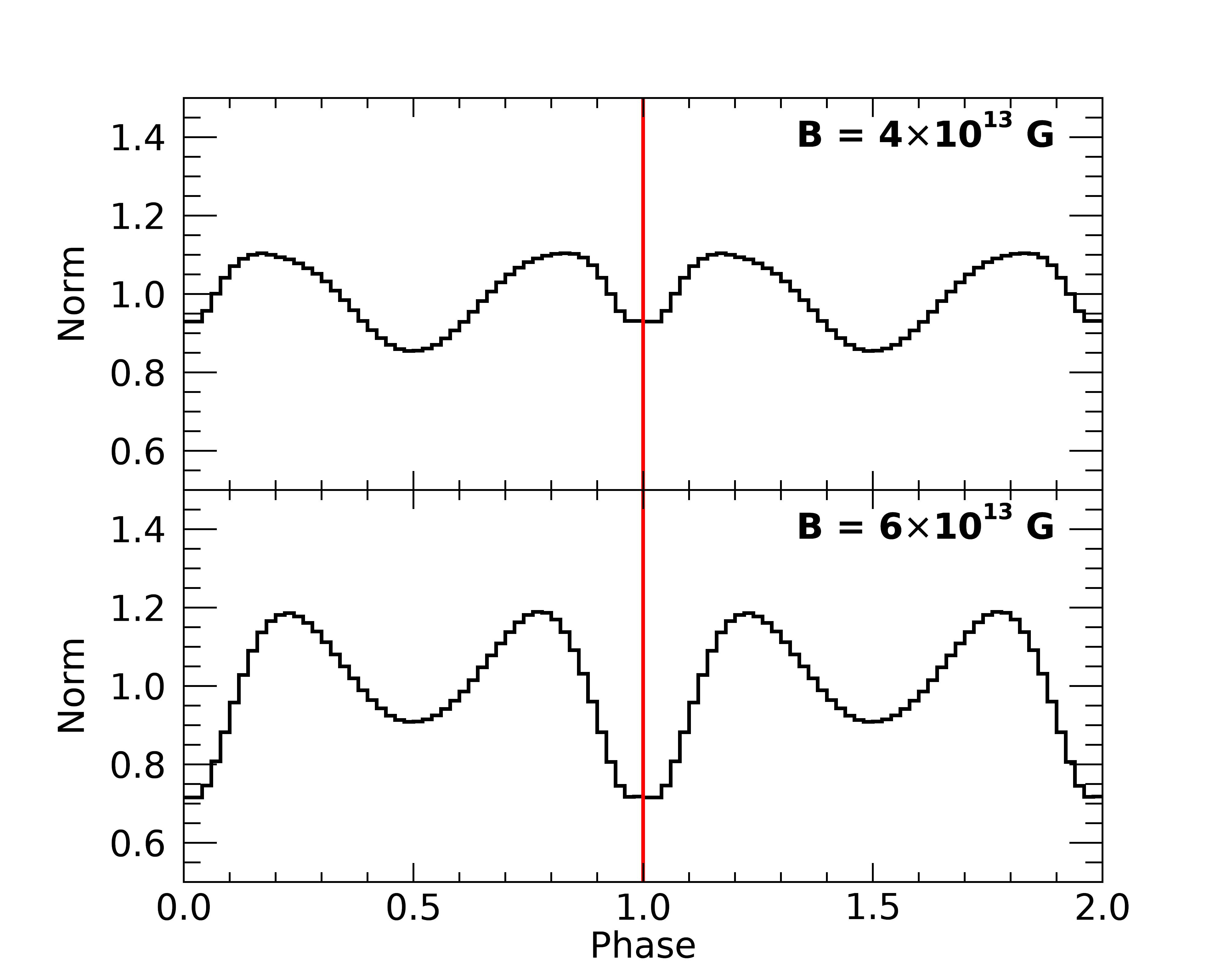}
    \caption{Pulse profiles in the \band{0.15}{1.5} range in the case of emission from a hydrogen atmosphere model at two point-like polar caps with $T_{\rm eff}=0.5$ MK and $B=4\times10^{13}$ G (upper panel) and $B=6\times10^{13}$ G (lower panel). We assumed $\xi=30\deg$, $\chi=35\deg$ and a compactness $M/R=0.1\msun/$km. The vertical red line shows the phase expected for the radio peak.}
    \label{fig:lcmod}
\end{figure}

Indeed, this model is oversimplified and unlikely to apply to the real case. 
Whatever the mechanism responsible for the surface emission, in fact, the presence of a strong magnetic field results in some degree of anisotropy in the emitted radiation. In the case of a magnetized atmosphere, more complicated energy-dependent beaming patterns are produced: they consist of a relatively narrow \textit{pencil-beam} aligned with the magnetic field, surrounded by a broader \textit{fan-beam} at intermediate angles and accounting for most of the escaping radiation (see e.g. \citealt{pav94}). 
The angular pattern of the emerging intensity depends also on the local surface temperature and magnetic field, so that the morphology of the pulse profiles can be extremely variegate. 
Using a partially ionized hydrogen atmosphere model \citep{sul09} with improved opacities from \citet{pot14}, we computed the expected pulse profiles, as described in \citet{rig19}. The best match with the data was obtained assuming emission from two antipodal hot spots with an effective temperature of $0.5$ MK, and $\xi=30\deg$, $\chi=35\deg$. In Fig. \ref{fig:lcmod} we show two examples with representative values of the magnetic field, $B=4\times10^{13}$ G and $B=6\times10^{13}$ G.
Although these pulse profiles qualitatively resemble that observed in \psrj, we note that they have been computed considering only the X-ray emission from the polar caps. The addition of a contribution from an extended part of the star surface would reduce the pulsed fractions of the light curves shown in Fig. \ref{fig:lcmod}.

\subsection{Connections with the XDINSs}

\setlength{\tabcolsep}{0.9em}
\begin{table*}[htbp!]
\centering \caption{Comparison between \psrj and the XDINSs}  
\label{tab:xdins}

\begin{tabular}{lccccccccc}
\toprule
Source						& $P$   & $\pdot$               & $B_{\rm p}$   & $E_{\rm cyc}$ & $B_{\rm cyc}$ & $L_X/\edot$   & Pulse	& PF    & Refs. \\[5pt]
RX							& s		& $10^{-14}$ s s$^{-1}$ & $10^{13}$ G   & eV            & $10^{13}$ G   &               &       & \%	&		\\[5pt]
\midrule\\[-5pt]
J0420.0$-$5022				& 3.45	& 2.76	& 2.0   & \dots	            & \dots                 & $0.31-0.38$   & single	& 13			& (1)	\\[5pt]
J0720.4$-$3125				& 16.78	& 18.6	& 11.3  & $254_{-30}^{+25}$	& $3.4_{-0.4}^{+0.3}$   & $99-157$      & double	& 11			& (2)	\\[5pt]
J0806.4$-$4123				& 11.37	& 5.6	& 5.1   & $241_{-12}^{+11}$	& $3.2_{-0.2}^{+0.2}$   & $10.6-16.7$   & single	& 6				& (1)	\\[5pt]
J1308.6$+$2127 (RBS 1223)   & 10.31	& 11.2	& 6.9   & $390_{-6}^{+6}$   & $5.16_{-0.08}^{+0.08}$& $31.5-39.6$   & double	& 18			& (3)	\\[5pt]
J1605.3$+$3249 (RBS 1556) 	& \dots	& \dots	& \dots & $353_{-48}^{+19}$	& $4.7_{-0.6}^{+0.3}$   & \dots         & \dots     & $<\,$1.4		& (4)	\\[5pt]
J1856.5$-$3754				& 7.06	& 2.98	& 2.9   & \dots             & \dots                 & $9.6-15.2$    & single	& 1.2			& (5)	\\[5pt]
J2143.0$+$0654 (RBS 1774) 	& 9.43	& 4.1	& 4.0   & $326_{-79}^{+56}$	& $4.3_{-1.0}^{+0.7}$   & $33.2-41.8$   & single	& 4				& (6-7) \\[5pt]

\midrule\\[-5pt]
\psrj	                	& 3.44  & 29.3  & 6.4  & $390_{-20}^{+10}$  & $5.2_{-0.3}^{+0.1}$   & $1.1-3.0$     & double    & 30            & (8)   \\[5pt]
\bottomrule\\[-5pt]
\end{tabular}

\raggedright
$B_{\rm p}$ and $B_{\rm cyc}$ are the magnetic field at the poles evaluated from the timing parameter and from the cyclotron energy, respectively. $E_{\rm cyc}$ values are taken from \citet{yon19}, while $L_X$ values from \citet{vig13}.\\
Specific references: (1) \citet{hab04}; (2) \citet{ham17}; (3) \citet{ham11}; (4) \citet{pir19}; (5) \citet{tie07}; (6) \citet{zan05}; (7) \citet{mig11} (8) This paper.

\end{table*}

\begin{figure}[htbp!]
    \centering
    \includegraphics[width=0.9\linewidth]{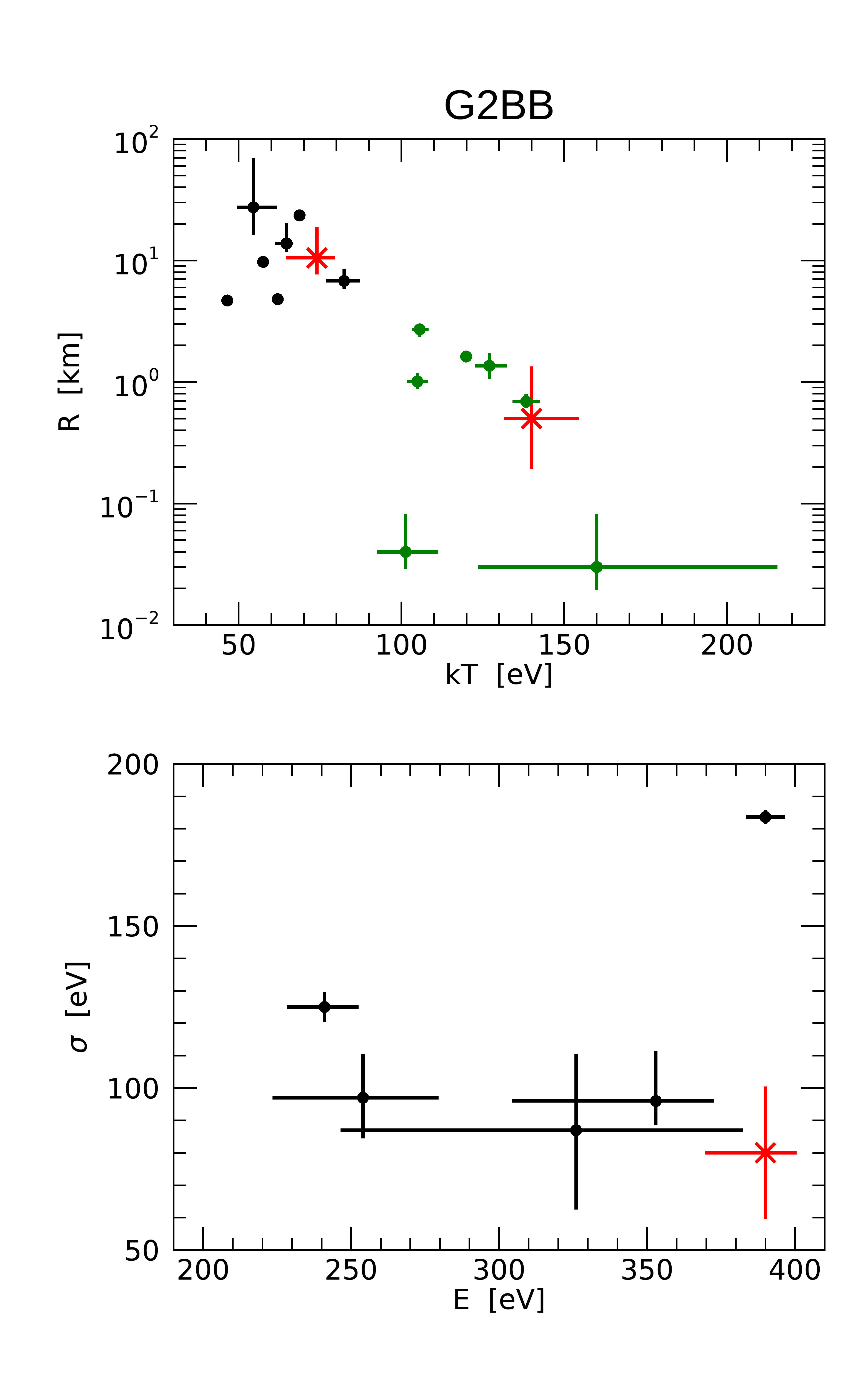}
    \caption{Comparison of the spectral parameters of the XDINSs (from \citealt{yon19}) and \psrj (red cross) obtained with two blackodies and a gaussian absorption line model (G2BB). The upper panel shows the blackbody radii (black: cold; green: hot) for the XDINSs and for \psrj (red cross). The lower panel illustrates the line width vs the line centroid energy (black dots are for XDINSs and the red cross for \psrj).}
    \label{fig:specall}
\end{figure}

Our spectral results, and in particular the presence of a broad absorption line, strengthen the similarity between \psrj and the XDINSs, for which similar spectral features have been reported (see Table \ref{tab:xdins}).
As it is illustrated in Fig. \ref{fig:specall}, not only the line properties, but also the best fit parameters of the continuum model are very similar to those recently reported in a systematic analysis of all the XDINS spectra with the G2BB model \citep{yon19}.

\begin{figure}[htbp!]
    \centering
        \includegraphics[width=1\linewidth]{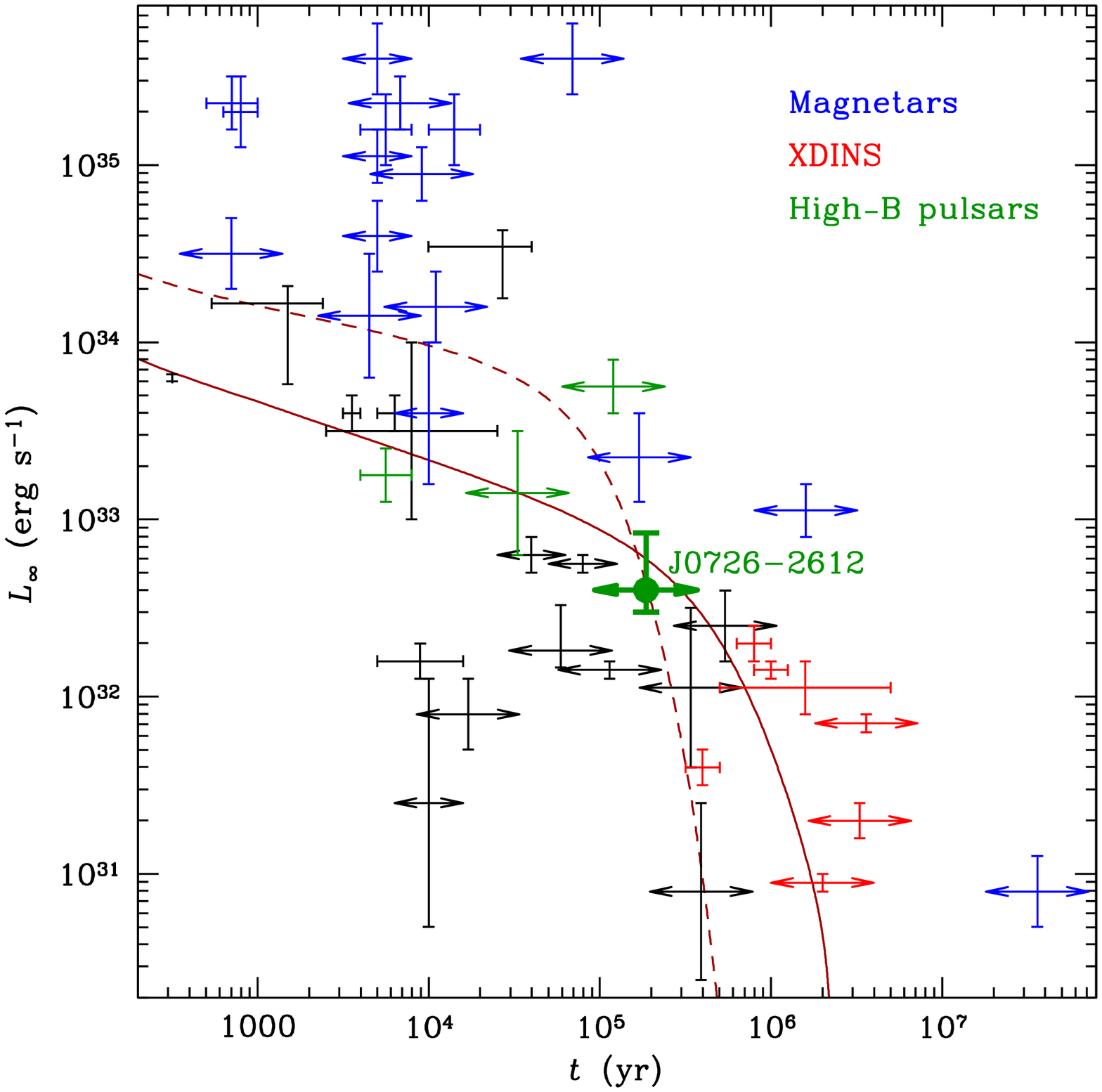}\\
    \caption{Thermal luminosities versus ages of isolated neutron stars. The same color coding of Fig. \ref{fig:p0p1} is used (in addition to \psrj, the three High-B pulsars are J1119$-$6127, J1718$-$3718 and J1819$-$1458). The solid and dashed lines are the theoretical cooling curves of a neutron star with mass $M=1.2\msun$ and the dipole magnetic field of \psrj ($B_p=6\times10^{13}$~G), with the heat blanketing outer envelope composed either of iron (solid line) or of accreted light elements (dashed line).}
    \label{fig:cooling}
\end{figure}

Considerations on the age-luminosity diagram shown in Fig.~\ref{fig:cooling} give even more strength to this analogy. The figure represents the bolometric luminosity of thermally emitting neutron stars against their ages, characteristic or kinematic. The luminosity of \psrj $L_\infty = (4.0_{-1.0}^{+4.4})\times10^{32}$ \lum corresponds to the cold component of the G2BB fit to the phase-averaged spectrum (for $d=1$ kpc).
This component is in fact representative of the cooling emission from the entire star surface (the inclusion of the hot component would not significantly change the result, adding only about 3\% to the total luminosity, well within the uncertainties). The observational data for other neutron stars are displayed in Fig. \ref{fig:cooling} as in \citet{pot18}; most of them are taken from \citet{vig13}, with some updates and additions.
The horizontal error bars show the uncertainties of kinematic ages, when available, otherwise the bars are replaced by arrows. 

The position of \psrj in this diagram is indeed close to the group of XDINSs. Its place can be considered as intermediate between the regions occupied by ordinary neutron stars, which have either smaller luminosities or smaller ages, magnetars, which generally have larger luminosities, and XDINSs, which have somewhat smaller luminosities and larger ages. For comparison we plot two cooling curves, with heavy (nonaccreted) and light (accreted) chemical elements in the outer heat-blanketing envelope. The cooling curves are calculated for a neutron star of mass $M=1.2\,M_\odot$ and the dipole magnetic field inferred for \psrj ($B_p = 6 \times 10^{13}$~G) using the code of \citet{pot18} with the equation of state BSk24 \citep{pea18}, singlet pairing type superfluidity of neutrons and protons (according to \citealt{mar08} and \citealt{bal07}, respectively, both in the parametrized form of \citealt{ho15}). The triplet pairing type superfluidity of neutrons is not included, because it is strongly suppressed by many-particle correlations, according to recent results of \citet{din16}. The latter suppression delays the onset of the Cooper pair breaking-formation mechanism of neutrino emission in the core of the neutron star and thus slows down the cooling, making the theoretical cooling curves compatible with the XDINS observations even without additional internal heating, which otherwise would be needed (e.g., \citealt{vig13}).

While most of the XDINSs have single-peaked pulse profiles, two of them (RX J1308.6$+$2127, \citealt{ham11}, and RX J0720.4$-$3125, \citealt{ham17}) show double-peaked profiles similar to \psrj, although with smaller pulsed fractions ($18 \%$ and $11 \%$, respectively). 
The remarkable difference between \psrj and these two XDINSs is the presence of radio emission in the former. Here we discuss the possibility that this is due an unfavourable orientation of their radio beam. Based on the radio beaming fraction of long period pulsars, \citet{kon09} estimated that one should observe a much larger number of XDINSs ($\sim$40) to  detect one with the radio beam crossing our line of sight.

\begin{figure}[htbp!]
    \centering
     \includegraphics[width=1\linewidth]{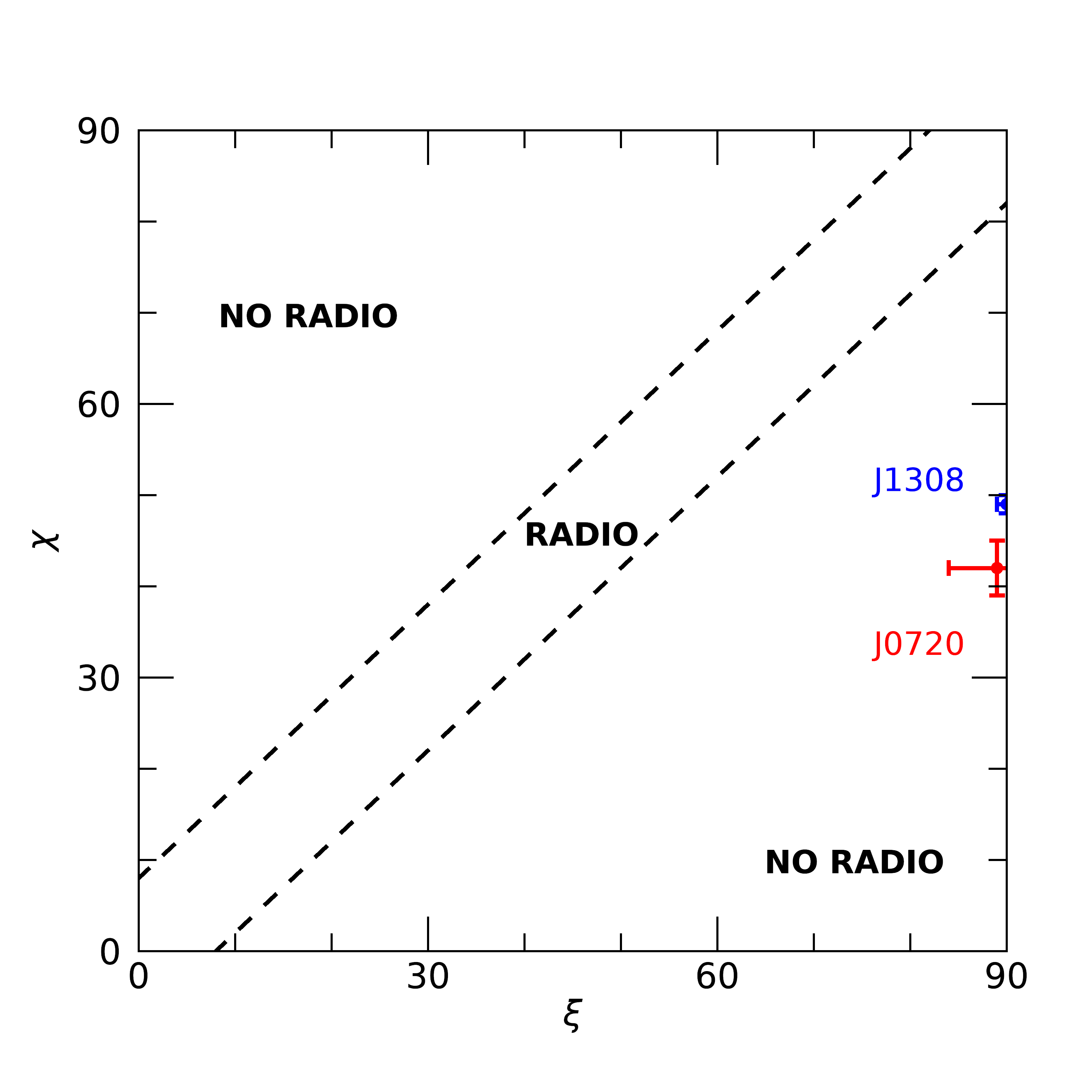}
    \caption{Visibilty of a radio beam with aperture of $\approx 8\deg$ as a function of the $\xi$ and $\chi$ angles. The estimated positions for RX J1308.6$+$2127 \citep[ blue dot]{ham11} and RX J0720.4$-$3125 \citep[ red dot]{ham17} are shown. }
    \label{fig:chixi}
\end{figure}

We have marked in Fig. \ref{fig:chixi} the values of the angles $\xi$ and $\chi$  estimated for RX J1308.6$+$2127 and RX J0720.4$-$3125 by \citet{ham11} and \citet{ham17}. 
They imply that these two pulsars  are nearly orthogonal rotators ($\xi\approx 90\deg$) seen with a large impact parameter $\eta=\abs{\chi-\xi}\approx 45\deg$.  
With the usual assumption that the radio beam coincides, or is close to, the magnetic dipole axis, such a large impact parameter can naturally account for the fact that their radio emission is not visible from the Earth. 
As an example, the dashed lines in Fig. \ref{fig:chixi} indicate the region where $\xi\sim\chi$ for which a radio beam with aperture of $\approx 8\deg$ would be visible. Contrary to the two XDINSs, \psrj should lie inside this region. Our atmosphere model used to compute the pulse profiles of Fig~\ref{fig:lcmod}, predicts that the radio pulse, that appears when the magnetic axis is in the plane defined by the line of sight and rotation axis, is at the phase of one of the two minima of the X-ray profile.  
Considering the current relative error in the radio and X-ray phase alignment (see Fig. \ref{fig:lcall}), this possibility cannot be excluded.

\section{Conclusions}

Our analysis of \xmm data of the slow, highly magnetized radio pulsar \psrj revealed the presence of a broad absorption line in its soft thermal spectrum, with parameters similar to those of the lines seen in most of the XDINSs. The X-ray pulse profile of \psrj is double-peaked and moderately energy-dependent. 
These findings reinforce the similarity between this radio pulsar and the XDINSs.
Assuming a distance of 1 kpc, the luminosity of \psrj is $L_{\infty} = (4.0_{-1.0}^{+4.4}) \times 10^{32}$ \lum. This is greater than its spin-down luminosity, as for the XDINSs (see Table \ref{tab:xdins}), but it is in reasonable agreement with the expected thermal luminosity of a $\approx 200$ kyr old pulsar (see Fig. \ref{fig:cooling}).

More observations are needed to reduce the uncertainty in the radio and X-ray phase alignment and  better constrain the geometry of \psrj. This can help to understand if  the detection of radio emission in this pulsar, and not in the XDINSs with a similar double-peaked X-ray pulse profile, is due only to orientation effects.

\begin{acknowledgements}
We are grateful to an anonymous referee for constructive suggestions. 
We acknowledge financial contribution from the agreement ASI-INAF n.2017-14-H.0. Part of this work has been funded using resources from the research grant “iPeska” (P.I. Andrea Possenti) funded under the INAF national call Prin-SKA/CTA approved with the Presidential Decree 70/2016.
This work is based on observations obtained with \xmm, an European Space Agency (ESA) science mission with instruments and contributions directly funded by ESA Member States and NASA. 
The work of A.Y.P. was supported by DFG and RFBR within the research project 19-52-12013.
The work of V.S. was supported by Deutsche Forschungsgemeinschaft (DFG, (grant WE 1312/51-1) and by the subsidy allocated to Kazan Federal University for the state assignment in the sphere of scientific activities (3.9780.2017/8.9).

\end{acknowledgements}

\bibliographystyle{aa}
\bibliography{bibliography}

\end{document}